\documentclass[a4paper,fleqn]{cas-dc}

\usepackage[numbers]{natbib}
\usepackage{lscape} 
\usepackage{amssymb}
\usepackage{amsmath}
\usepackage{latexsym}
\usepackage{bigstrut}
\usepackage{soul}
\usepackage{color, xcolor}
\usepackage{ulem}
\usepackage{placeins} 

\usepackage{hyperref}
\hypersetup{
	colorlinks=true,
	linkcolor=cyan,
	filecolor=cyan,
	urlcolor=cyan,
	citecolor=cyan,
}
\usepackage{cleveref}
\usepackage{algorithm}  
\usepackage{algpseudocode}

\begin{document}
\let\WriteBookmarks\relax
\def\floatpagepagefraction{1}
\def\textpagefraction{.001}

\shorttitle{Automatic diagnosis of cardiac magnetic resonance images based on semi-supervised learning}    

\shortauthors{Hejun Huang et al.}  

\title [mode = title]{Automatic diagnosis of cardiac magnetic resonance images based on semi-supervised learning}  



%

\author[1]{Hejun Huang}

\author[1,2]{Zuguo Chen}
\cormark[1]
\cortext[1]{Corresponding author. School of Information and Electrical Engineering, Hunan University of Science and Technology, Xiangtan, 411201, China}
\ead{zg.chen@hnust.edu.cn}

\author[1]{Yi Huang}

\author[3]{Guangqiang Luo}

\author[1]{Chaoyang Chen}

\author[4]{Youzhi Song}

\affiliation[1]{organization={School of Information and Electrical Engineering, Hunan University of Science and Technology},
	city={Xiangtan},
	postcode={411201}, 
	country={China}}
\affiliation[2]{organization={Shenzhen Institute of Advanced Technology, Chinese Academy of Sciences},
	city={Shenzhen},
	postcode={518055}, 
	country={China}}
\affiliation[3]{organization={Guilin Medical University},
	city={Guilin},
	postcode={541001}, 
	country={China}}

\affiliation[4]{organization={Rucheng County Hospital of Traditional Chinese Medicine},
	city={Chenzhou},
	postcode={424100}, 
	country={China}}

\begin{abstract}
Cardiac magnetic resonance imaging (MRI) is a pivotal tool for assessing cardiac function. Precise segmentation of cardiac structures is imperative for accurate cardiac functional evaluation. This paper introduces a semi-supervised model for automatic segmentation of cardiac images and auxiliary diagnosis. By harnessing cardiac MRI images and necessitating only a small portion of annotated image data, the model achieves fully automated, high-precision segmentation of cardiac images, extraction of features, calculation of clinical indices, and prediction of diseases. The provided segmentation results, clinical indices, and prediction outcomes can aid physicians in diagnosis, thereby serving as auxiliary diagnostic tools. Experimental results showcase that this semi-supervised model for automatic segmentation of cardiac images and auxiliary diagnosis attains high accuracy in segmentation and correctness in prediction, demonstrating substantial practical guidance and application value.
\end{abstract}



\maketitle

\section{Introduction}
\label{sec1}
Cardiovascular diseases stand as a primary cause of mortality globally, imposing a significant burden on public health worldwide \cite{Ribeiro_Left_Ventricle}. Cardiac functional analysis constitutes a crucial aspect of cardiovascular disease diagnosis. Cardiac magnetic resonance imaging, offering high-resolution images without radiation, is increasingly becoming a common modality in cardiac diagnostics, heralded as the "gold standard" for cardiac disease diagnosis. CMR imaging necessitates the acquisition of images spanning at least one cardiac cycle, which comprises diastole and systole phases. By integrating images from end-diastole and end-systole, the segmentation of the left ventricle, right ventricle, and myocardium enables the quantitative assessment of ventricular volumes, myocardial mass, ejection fraction, and other clinical indices. These clinical indices serve as classification features for cardiac diseases in prognostic applications. However, the accuracy of clinical index calculations relies heavily on the precision of cardiac structure segmentation.

Traditional methods for cardiac image segmentation can be categorized into those based on prior knowledge. Weak prior knowledge methods, such as thresholding, Gaussian modeling, and level set methods, require interaction with experts for completion and often yield subpar segmentation results. Strong prior knowledge methods utilize statistical models to incorporate prior knowledge as constraints on the final segmentation results, including shape-based deformable models \cite{Zhao_Congenital_aortic}, models based on cardiac motion and appearance \cite{Mitchell_3D_activate}, and atlas-based methods \cite{Bai_Muti-atlas_segmentation}. However, due to the stringent imposition of foundational information like cardiac shape, strong prior knowledge methods are prone to overfitting and may exhibit poor segmentation performance on data outside the predefined range of foundational information.

Algorithms based on deep learning excel in automatically extracting complex features from data for segmentation purposes. Represented by U-Net \cite{Ronneberger_UNet}, methods based on Convolutional Neural Networks (CNNs) have been widely utilized in medical image segmentation since their inception \cite{Siddique_unet_review}. Vesal et al. \cite{Vesal_Dilated_convolutions} introduced a 3D U-Net that combines local and global information, leveraging dilated convolutions to expand the receptive field of the network's lowest layers. Chen et al. \cite{Chen_Mutiview_two-task} proposed a multi-view dual-task recursive attention U-Net, where recursive structures and dilated convolutions extract multi-view features while attention mechanisms precisely delineate the left atrium and depict atrial scars. In recent years, Vision Transformers have garnered significant attention due to their ability to encode long-range dependencies. However, Vision Transformers lack robust local inductive bias and typically require extensive training on large datasets to achieve satisfactory results. Consequently, in the field of medical image segmentation, much of the work involves integrating CNNs with Vision Transformers. Zhang et al. \cite{Zhang_Transfuse} merged Transformers and CNNs in parallel to enhance the efficiency of global information processing. Similarly, Ji et al. \cite{Ji_Muti-compound} utilized transformers to model rich contextual dependencies and semantic relationships for precise biomedical image segmentation. Wang et al. \cite{Wang_Transbts} introduced Transformers into the bottleneck layers of a CNN encoder-decoder structure, exploiting the powerful local information learning capability of CNNs and the long-distance feature dependency modeling capability of Transformers, cleverly combining the strengths of both networks to unleash their full potential.

The attention mechanism enables networks to adaptively focus on crucial components, effectively enhancing the performance of segmentation networks. For segmentation tasks, attention mechanisms are divided into channel attention and spatial attention. Channel attention directs focus towards significant objects \cite{Wang_ECA-Net,Chen_Channel-Unet,Hu_SE}, while spatial attention highlights important regions \cite{Guo_Sa-unet,Jaderberg_Spatial_transformer}. To address class imbalance issues in left atrium segmentation, Cui et al. \cite{Cui_Mutiscale_attention} devised a loss function and incorporated a multiscale input pyramid into the attention network to better represent intermediate features, while employing deep supervision output layers to aid the network in handling shape variability. Sinha et al. \cite{Sinha_Muti-scale_self-guided} introduced a guided self-attention mechanism to capture richer contextual dependencies. PraNet \cite{Fan_PraNet} proposed reverse attention to enhance segmentation boundary accuracy but lacked adequate attention to important regions. To mitigate the computational complexity of self-attention, SegNeXt \cite{Guo_SegNeXt} proposed a more efficient multiscale convolutional attention mechanism as a substitute, yielding favorable results. While previous methods focused solely on spatially significant regions, they lacked attention to important objects in the channel dimension. CBAM \cite{Woo_cbam} integrated channel and spatial attention, yet spatial attention maps were derived through channel compression, resulting in uniform spatial attention weight distribution across channels. Hence, enhancing the adaptive capability of attention mechanisms by dynamically distributing attention weights in both channel and spatial dimensions can effectively improve the performance of segmentation networks.

Deep learning-based methods have yielded promising results in automatic segmentation of cardiac images. To mitigate reliance on large annotated datasets, semi-supervised learning-based approaches for cardiac image segmentation have emerged as a new research direction. Combining supervised and unsupervised learning, semi-supervised learning autonomously leverages unlabeled data to enhance learning performance \cite{Yang_A_survey}. Given the high cost and difficulty of obtaining labeled data for segmentation tasks, semi-supervised segmentation has gained traction, aiming to extract useful information from unlabeled data as much as possible. Chen et al. \cite{Dong_Tri-net2018} constructed three models with branches to generate pseudo-labels through a voting mechanism, yielding satisfactory results particularly when abundant unlabeled data are available, despite introducing noise. Yu et al. \cite{Yu_Uncertainty-aware2019} utilized Monte Carlo sampling to obtain uncertainty maps from a teacher model to guide a student model in progressively acquiring reliable information. Li et al. \cite{Li_Shape2020} introduced signed distance map regression with shape and position priors and employed a discriminator as a regularization term to enhance segmentation stability. Semi-supervised segmentation extracts effective and reliable information from unlabeled data to improve model performance, reducing dependency on segmentation labels to a certain extent. By combining effective methods and designing rational semi-supervised segmentation architectures, the challenge of limited labeled samples in cardiac image annotation can be mitigated, yielding reliable cardiac image segmentation results.

Methods for automatic cardiac diagnosis based on cardiac images primarily involve automatic computation of relevant clinical indices from the images followed by disease classification. Zhao et al. \cite{Zhao_Congenital_aortic} computed modal indices of aortic shape and motion and utilized a support vector machine (SVM) classifier to differentiate between normal subjects and those with connective tissue disorders. Suinesiaputra et al. \cite{Suinesiaputra_Statistical} employed statistical shape modeling of the left ventricle for myocardial infarction classification. However, these approaches utilize fewer physiological indicators and suffer from severe limitations in distinguishing disease categories. Khened et al. \cite{Khened_Fully_convolutional} utilized deep learning methods for automatic cardiac structure segmentation, extracting relevant features for training a two-stage ensemble classifier for automatic cardiac disease diagnosis, achieving significant breakthroughs. Nonetheless, challenges persist regarding segmentation accuracy and excessive reliance on unlabeled data for segmentation methods.

This article proposes a semi-supervised cardiac image automatic segmentation and auxiliary diagnosis model by combining channel prior-based convolutional attention \cite{Huang_CPCA} and semi-supervised segmentation structures based on complementary consistency \cite{Huang_CCNet}. Leveraging a small amount of labeled data, the model achieves high-quality segmentation results, followed by volume reconstruction and evaluation of cardiac structures, thereby computing clinical indices suitable for disease discrimination. Additionally, a two-layer ensemble classifier is designed using machine learning classifiers. The ensemble classifier serves as a regularization and feature importance balancing mechanism, further optimizing classification accuracy. Consequently, the contributions and novelty of this paper are summarised as follows:
\begin{itemize}
	\item The proposal of a dual-layer ensemble classifier enhances the classifier's generalization capability while improving classification accuracy.
	\item The proposed semi-supervised cardiac image automatic segmentation and auxiliary diagnosis model requires only a small amount of annotated image data and can achieve fully automated, high-precision cardiac image segmentation, clinical index calculation, feature extraction for classification, and disease prediction.
\end{itemize}

\section{Methods}
\label{sec2}

\begin{figure*}[htb]
	\centering
	\includegraphics[scale=.5]{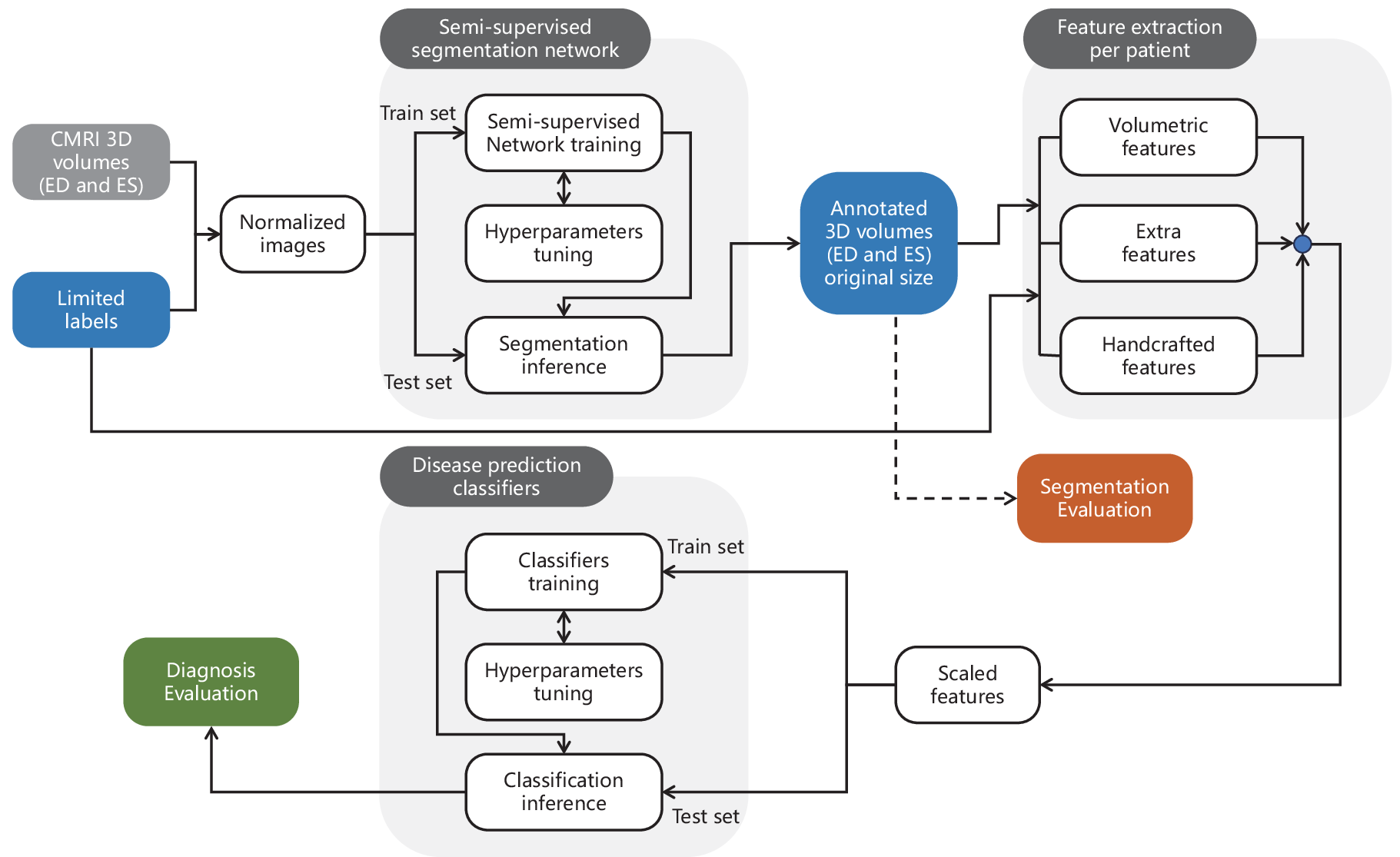}
	\caption{A schematic depiction of the semi-supervised automatic segmentation and auxiliary diagnostic process is presented. CMR 3D images and partial labels act as inputs, processed through a semi-supervised segmentation network to generate segmentation results. Subsequently, these segmentation outcomes are transformed into various features and supplied to a disease prediction classifier, culminating in diagnostic outcomes.}
	\label{fig1}
\end{figure*}

\subsection{Network overall architecture}
\label{subsec2-1}
The model architecture summarized in this paper is depicted in Figure \ref{fig1}. Cardiac magnetic resonance images (comprising end-diastolic and end-systolic phases) and partially annotated labels serve as inputs. Input images undergo image standardization to eliminate distribution differences between images, facilitating rapid convergence during network training. Preprocessed images are divided into training and testing sets and fed into the semi-supervised segmentation network. The training set is utilized for network training and cross-validation, with hyperparameters fine-tuned based on validation results to achieve optimal performance. The optimal model from the training phase is used for segmentation inference on the testing set, yielding annotated CMR images. The segmentation results are then inputted into the feature extraction module for case-by-case feature extraction. Extracted features include voxel features reconstructed from cardiac structures, additional features provided by the dataset, and manually designed derived features. All features, after feature standardization, are divided into training and testing sets and fed into the disease prediction classifier. The training set is employed for classifier training and cross-validation, with hyperparameters fine-tuned based on validation results to achieve optimal performance. The optimal model from the training phase is utilized for classification inference on the testing set, resulting in disease predictions. Finally, segmentation evaluation and diagnostic assessment can be performed based on the segmentation and classification results.

\subsection{Semi-supervised automatic cardiac image segmentation network}
\label{subsec2-2}
\begin{figure*}[htb]
	\centering
	\includegraphics[scale=.55]{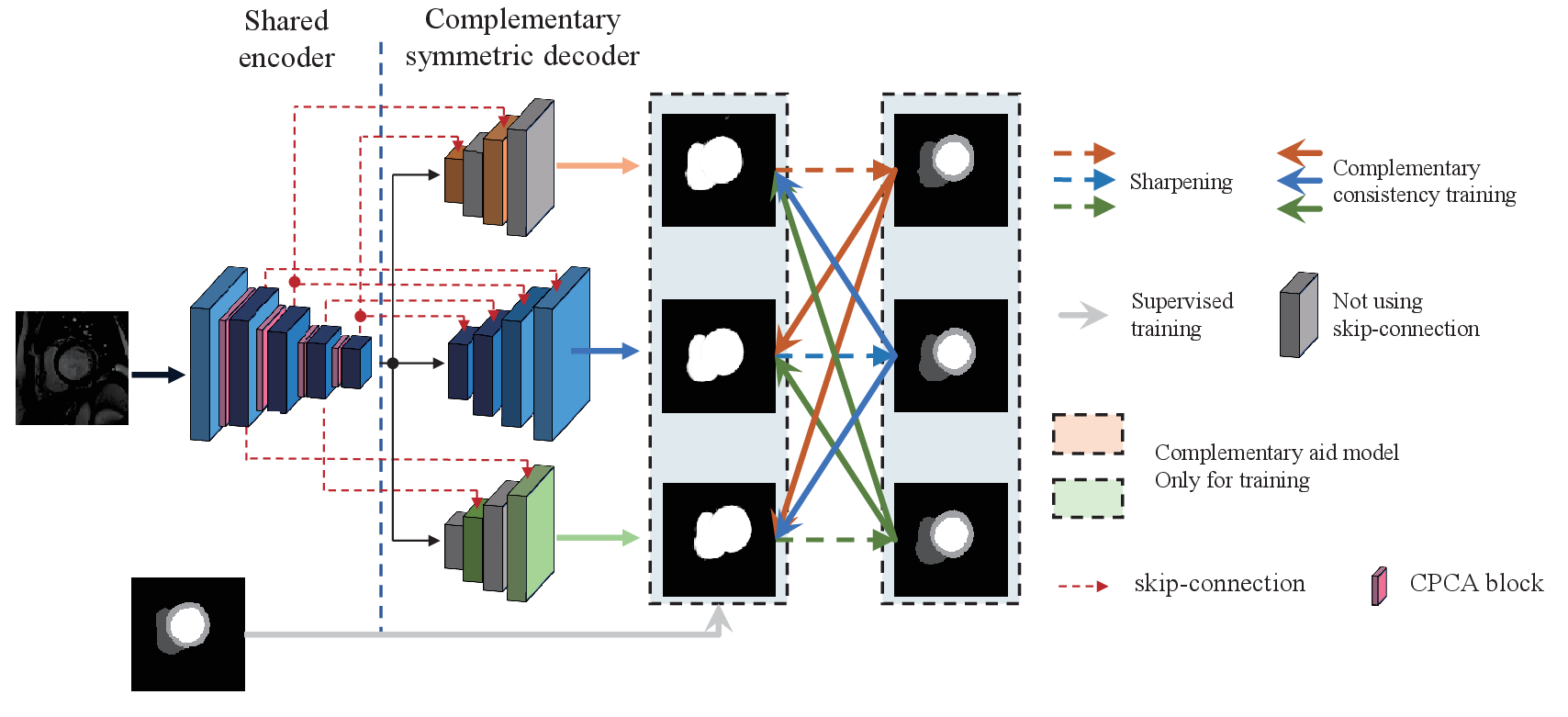}
	\caption{A schematic diagram of the semi-supervised cardiac image segmentation network structure is depicted. The network comprises a shared encoder with a CPCA module and three decoders. Among them, two decoders, formed by alternately utilizing skip connections, constitute complementary auxiliary decoders. The probability maps generated by the decoders are sharpened to obtain pseudo-labels for mutual learning between the decoders.}
	\label{fig2}
\end{figure*}
The semi-supervised cardiac segmentation network structure proposed in this paper is depicted in Figure \ref{fig2}. Integrating the channel prior convolutional neural network into the complementary consistency semi-supervised segmentation architecture, utilizing a shared encoder to reduce bias during feature extraction, achieves efficient and accurate semi-supervised automatic segmentation.

The network is based on U-Net \cite{Ronneberger_UNet} and consists of an encoder and a decoder. The first four encoding layers of the encoder are followed by corresponding-sized CPCA modules. Additionally, two complementary auxiliary decoders are incorporated into the decoder part. These auxiliary decoders utilize skip connections alternately to form model-level perturbations, extracting complementary information from unlabeled data. The main decoder and two complementary auxiliary decoders share the encoder output. When labels are available for training images, probability maps generated by the three auxiliary models are compared with the ground truth labels to compute supervised loss, employing the commonly used Dice loss for segmentation. In the absence of labels for training images, probability maps generated by the three auxiliary models are sharpened to form pseudo-labels. The loss is computed by comparing pseudo-labels generated by each decoder with probability maps generated by the other two decoders, facilitating unsupervised learning, with unsupervised loss calculated using mean squared error (MSE) between pseudo-labels and probability maps.

\subsection{Clinical metrics calculation and categorical feature extraction}
\label{subsec2-3}
The subsequent stage following cardiac image segmentation involves the reconstruction of cardiac structures using the segmentation outcomes. This facilitates the quantitative computation of pertinent clinical indices. These indices, including their ratios and other derived data, can be utilized as features for the classification of heart diseases. The features employed for cardiac disease classification in this paper encompass volume features, additional features, and designed features.

Volume features encompass directly computed volumes of cardiac structures as well as features derived from volume calculations. End-diastolic left ventricular volume, end-systolic left ventricular volume, end-diastolic right ventricular volume, end-systolic right ventricular volume, and end-systolic myocardial volume are computed by multiplying the number of voxels corresponding to the respective cardiac structures in all MRI slices of the patient by the spatial resolution, as expressed by the following formula:
\begin{equation}
    V=\sum_i\sum_j\sum_kn\times s
    \label{eq1}
\end{equation}
where $n$ denotes the number of voxel points in the layer, and $s$ denotes the spatial resolution.

Ventricular ejection fraction (EF) is calculated using the formula:
\begin{equation}
    \mathrm{EF}=\frac{V_{_{ED}}-V_{_{ES}}}{V_{_{ED}}}\times100\%
    \label{eq2}
\end{equation}
where $V_{_{ED}}$ denotes end-diastolic ventricular volume and $V_{_{ES}}$ denotes end-systolic ventricular volume.

Myocardial mass ($M_{myo}$) is computed using the formula:
\begin{equation}
    M_{myo}=V_{myo}\times\rho_{myo}
    \label{eq3}
\end{equation}
where $V_{myo}$ denotes myocardial volume and $\rho_{myo}$ denotes myocardial density, with end-diastolic myocardial density often recognized as 1.05 g/mL.

Additional features encompass the height and weight of each patient provided in the dataset, as well as the calculated human body surface area (BSA). BSA is calculated using the formula:
\begin{equation}
    BAS=\sqrt{\frac{Height\times Weight}{3600}}
    \label{eq4}
\end{equation}
where $Height$ is the patient's height and $Weight$ is the patient's weight.

The design feature elucidates the variation in myocardial wall thickness (MWT). Employing the Canny edge detection algorithm for delineating the inner and outer contours of myocardial segmentation, let $I$ and $E$ represent the sets of pixels corresponding to the inner and outer contours, respectively. MWT is comprised of the collection of shortest Euclidean distances ($d$) from a pixel in the inner contour to any pixel in the outer contour. The computational formula of MWT in short-axis (SA) slices is as follows:
\begin{equation}
    MWT\mid{SA}=\left\{\min_{e\in E}d(i,e),i\in I\right\}
    \label{eq5}
\end{equation}
By computing the average or standard deviation of MWT for each short-axis slice and aggregating them in the long-axis (LA) direction, it is feasible to estimate the variation in myocardial wall thickness. Consequently, there are four features describing the variation in myocardial wall thickness at both end-diastole and end-systole. Table \ref{tab1} illustrates all cardiac disease classification features and their meanings employed in this study.

\subsection{Disease prediction classifier}
\label{subsec2-4}
\begin{figure*}[htb]
	\centering
	\includegraphics[scale=.6]{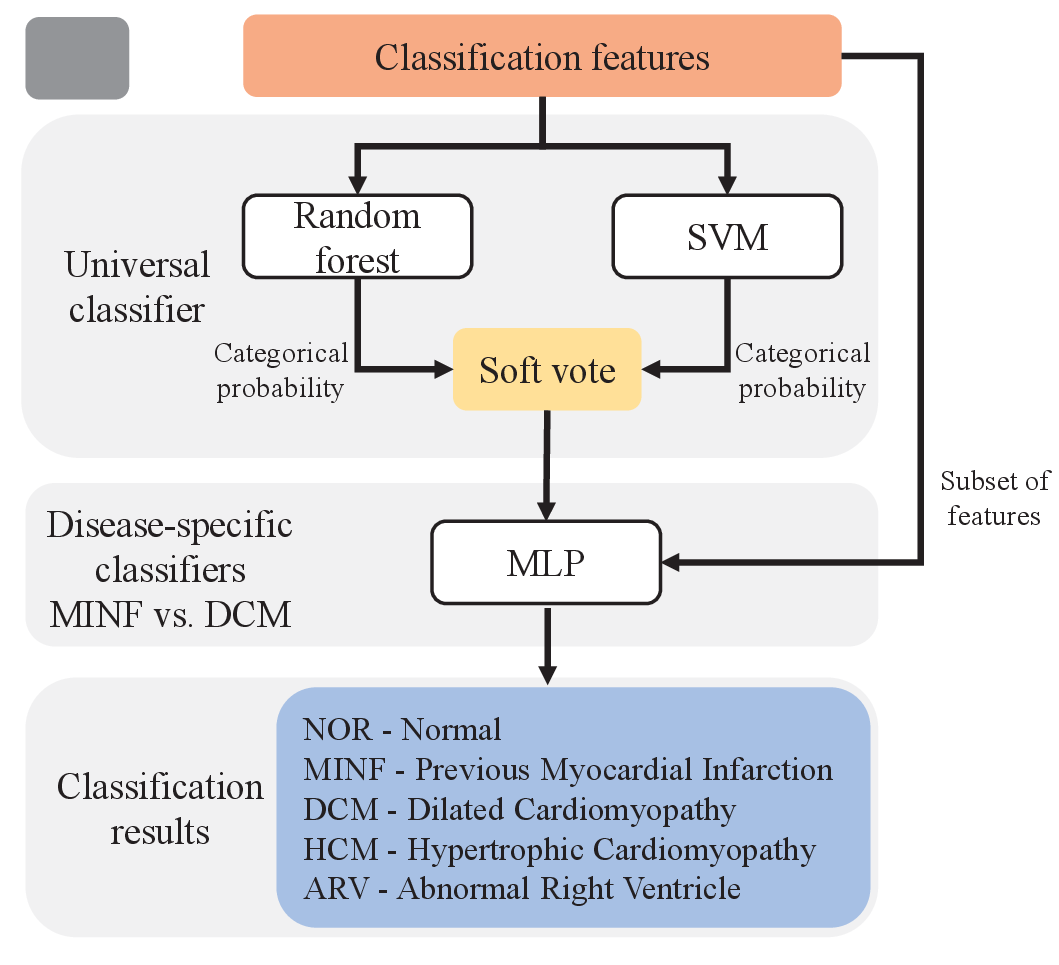}
	\caption{Diagram of the dual-layer ensemble classifier structure.}
	\label{fig3}
\end{figure*}
The paper proposes a dual-layer integrated classifier for disease classification and prediction, with the classifier structure depicted in Figure \ref{fig3}. This classifier consists of two layers. The first layer is a generic classifier, utilizing all classification features for classifying into five categories. The generic classifier comprises a combination of Random Forest and Support Vector Machine (SVM) through a soft voting mechanism. Soft voting entails independent prediction of all categories by all sub-classifiers, where the category probabilities predicted by all sub-classifiers for the sample are weighted and summed, and the category with the highest probability is selected as the predicted category for that sample. Random Forest is a decision tree-based classifier, while SVM is adept at classifying from high-dimensional features. The combination of these two classifiers aids in enhancing the model's generalization capability and increasing prediction accuracy. However, due to the symptoms of reduced myocardial contraction function present in both chronic myocardial infarction (MINF) and dilated cardiomyopathy (DCM), the generic classifier still exhibits inaccuracies in classifying certain samples.
\begin{table}[htbp]
	\centering
	\caption{Classification featrues of heart disease.}
	\begin{tabular}{lccc}
		\toprule
		\textbf{Feature} & \textbf{LV} & \textbf{RV} & \textbf{MYO} \\
		\midrule
		\multicolumn{4}{c}{\textbf{Volumetric features (Vol.)}} \\
		Vol.  & \checkmark     & \checkmark     & \checkmark \\
		Mass  &       &       & \checkmark \\
		Vol.(LV)/Vol.(RV) & \checkmark     & \checkmark     &  \\
		Vol.(MYO)/Vol.(LV) & \checkmark     &       & \checkmark \\
		EF    & \checkmark     & \checkmark     &  \\
		\multicolumn{4}{c}{\textbf{Myocardial wall thickness (MWT)}} \\
		max(mean(MWT|SA)|LA) &       &       & \checkmark \\
		stdev(mean(MWT|SA)|LA) &       &       & \checkmark \\
		mean(stdev(MWT|SA)|LA) &       &       & \checkmark \\
		stdev(stdev(MWT|SA)|LA) &       &       & \checkmark \\
		\multicolumn{4}{c}{\textbf{Extra features}} \\
		Height &       &       &  \\
		Weight &       &       &  \\
		BSA   &       &       &  \\
		\bottomrule
	\end{tabular}%
	\label{tab1}%
\end{table}%
To enhance the predictive accuracy of the classifier, this paper employs a Multi-Layer Perceptron (MLP) in the second layer for secondary classification of these two diseases. The primary distinctions between MINF and DCM include two key points: firstly, during myocardial contraction, MINF exhibits uneven myocardial wall thickness changes, whereas DCM's myocardial wall thickness is uniform; secondly, DCM is often associated with left ventricular dilation. The input features for the second-layer classifier are the distinguishing features between MINF and DCM, rather than all features. Additionally, MLP, utilizing fully connected neural networks, can learn high-dimensional information about classification features. This allows the disease-specific classifier based on MLP to further optimize the classification results for MINF and DCM.

\section{Experiments and results}
\label{sec3}

\begin{figure*}[p]
	\centering
	\includegraphics[scale=.95]{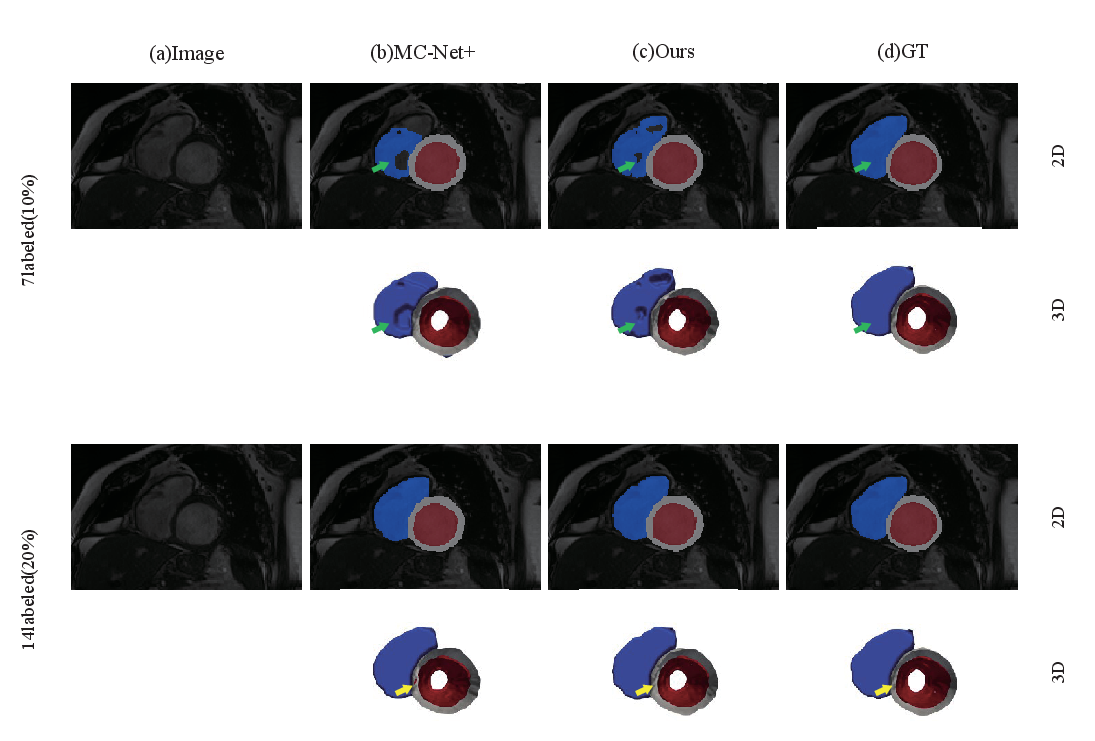}
	\caption{Visualization of segmentation results for comparison purposes. Blue represents the right ventricle (RV). White represents the left ventricular myocardium (MYO). Red represents the left ventricle (LV).}
	\label{fig4}
\end{figure*}

\begin{figure*}[p]
	\centering
	\includegraphics[scale=.95]{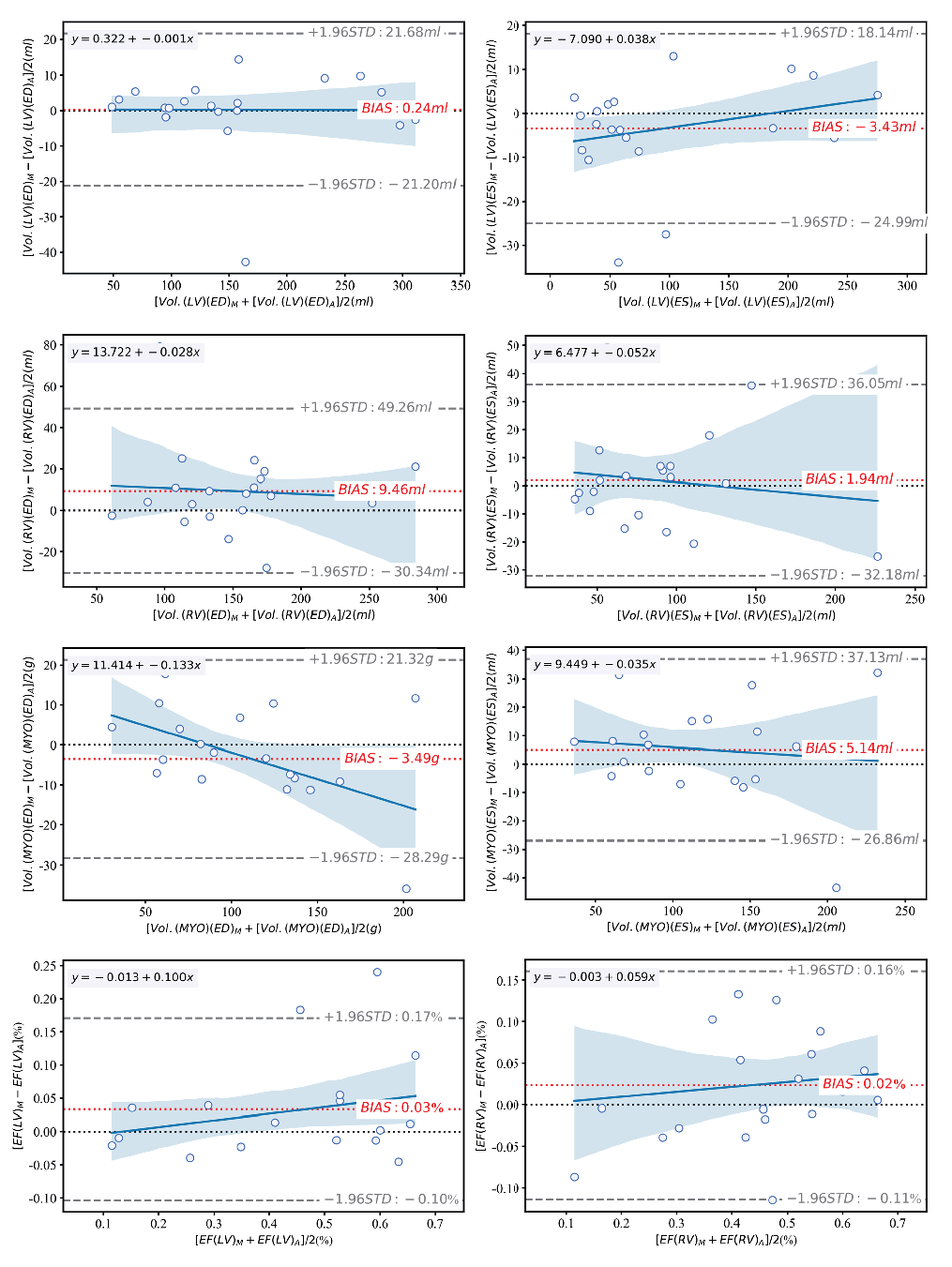}
	\caption{Error analysis using Bland-Altman plot and linear regression, with semi-supervised training on 7 annotated data (10\%).}
	\label{fig5}
\end{figure*}

\begin{figure*}[p]
	\centering
	\includegraphics[scale=.95]{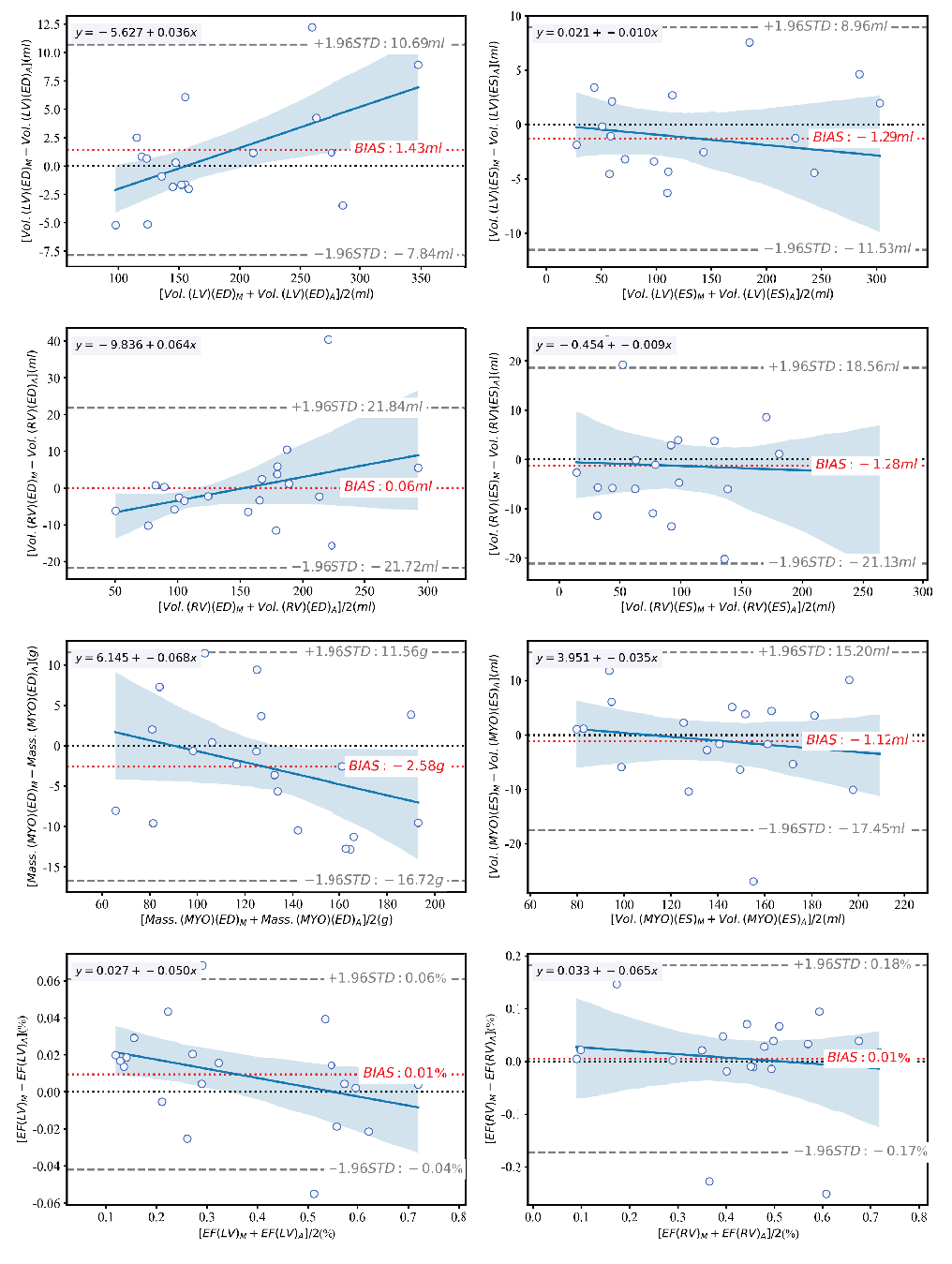}
	\caption{Error analysis using Bland-Altman plot and linear regression, with fully supervised training on 70 annotated data (100\%).}
	\label{fig6}
	
\end{figure*}

\begin{figure*}[p]
	\centering
	\includegraphics[scale=.75]{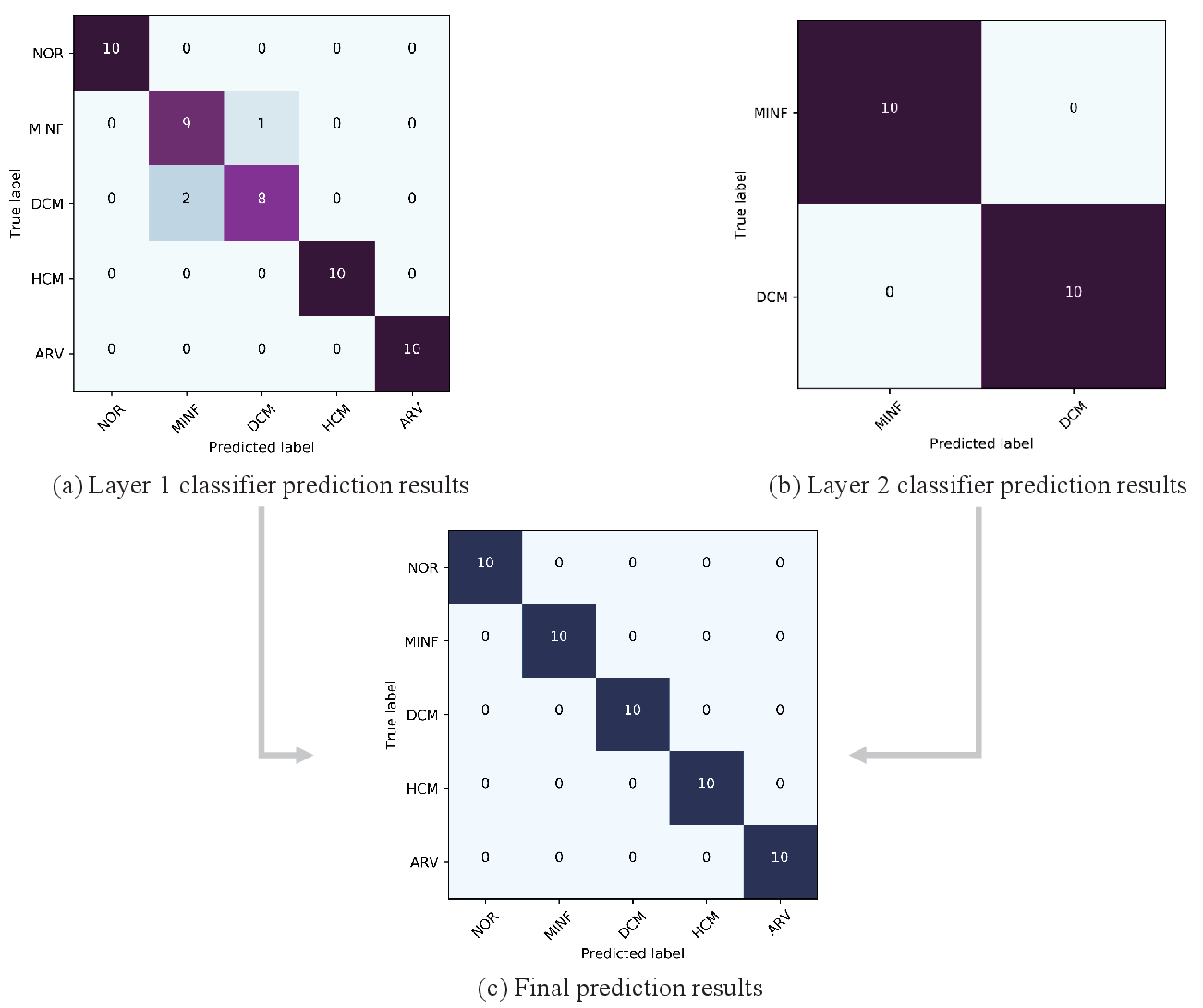}
	\caption{Confusion matrix for disease prediction.}
	\label{fig7}
\end{figure*}

\subsection{Dataset and implementing details}
\label{subsec3-1}
The ACDC dataset comprises samples from 150 different patients. Based on physiological parameters delineated in medical reports, the dataset consists of 5 categories: Normal (NOR), Previous Myocardial Infarction (MINF), Dilated Cardiomyopathy (DCM), Hypertrophic Cardiomyopathy (HCM), and Abnormal Right Ventricle (ARV), with 30 samples in each category. Each sample image spans from the base to the apex of the left ventricle, containing 28 to 40 short-axis slices, with a slice thickness ranging from 5 to 8 mm and slice gaps of 5 or 10 mm. The spatial resolution of the images ranges from 1.37 to 1.68 mm²/pixel. The organizers divided the dataset into two stages: 100 cases for training and 50 for testing. Manual annotations include left ventricle, right ventricle, and myocardium at end-diastole and end-systole.

In our experiments, the images from the training phase were split into training, validation, and test sets at a ratio of 7:1:2. The images from the testing phase were utilized for final evaluation. The experiments were conducted using Pytorch 1.9.1 + TensorFlow 1.13.1 + CUDA 11.1 framework and Python 3.6.5 on an NVIDIA RTX 3090 GPU. Evaluation metrics at the segmentation stage include Dice score \cite{Edlbode_Optimization} for internal segmentation assessment, Average Surface Distance (ASD), and 95\% Hausdorff Distance (95HD) for contour segmentation assessment. The formulas for Dice score, ASD, and HD calculations are provided in equations \ref{eq6}, \ref{eq7}, and \ref{eq8} respectively. Accuracy (ACC) is employed as the evaluation metric at the classification stage.

\begin{equation}
	\mathrm{Dice}(T,P)=\frac{2\sum_{i=1}^IT_iP_i}{\sum_{i=1}^IT_i+\sum_{i=1}^IP_i}
	\label{eq6}
\end{equation}
\begin{equation}
	\begin{aligned}
		\mathrm{ASD}(T^{\prime},P^{\prime})=&\frac{1}{\left|T^{\prime}\right|+\left|P^{\prime}\right|}\Bigg(\sum_{t\in T^{\prime}}\min_{p^{\prime}\in P^{\prime}}\left\|t^{\prime}-p^{\prime}\right\|+ \\ 
		&
		\sum_{p^{\prime}\in P^{\prime}}\min_{t^{\prime}\in T^{\prime}}\left\|p^{\prime}-t^{\prime}\right\|\Bigg)
		\label{eq7}
	\end{aligned}
\end{equation}
\begin{equation}
	\mathrm{HD}(T',P')=\max\left\{\max_{t'\in P'}\min_{p'\in T'}\|t'-p'\|,\max_{p'\in P'}\min_{t'\in T'}\|p'-t'\|\right\}
	\label{eq8}
\end{equation}

\subsection{Segmentation performance comparison experiment}
\label{subsec3-2}
\begin{table*}[htbp]
	\centering
	\caption{Comparison of quantitative results of different methods on the ACDC dataset (where data with * are from MC-Net+ \cite{Wu_Mutual_consistency2022}).}
	\begin{tabular}{lllllll}
		\toprule
		\multirow{2}[4]{*}{Methods} & \multicolumn{2}{l}{\#Scans used} &       & \multicolumn{3}{l}{Metrics} \\
		\cmidrule{2-3}\cmidrule{5-7}          & Labeled & Unlabeled &       & Dice(\%)↑ & 95HD(voxel)↓ & ASD(voxel)↓ \\
		\midrule
		U-Net* & 7(10\%) & 0     &       & 77.34 & 9.18  & 2.45 \\
		U-Net* & 14(20\%) & 0     &       & 85.15 & 6.20   & 2.12 \\
		U-Net* & 70(All) & 0     &       & 91.65 & 1.89  & 0.56 \\
		UA-MT \cite{Yu_Uncertainty-aware2019} (MICCAI) & 7(10\%) & 63(90\%) &       & 81.58 & 12.35 & 3.62 \\
		SASSNet \cite{Li_Shape2020} (MICCAI) & 7(10\%) & 63(90\%) &       & 84.14 & \textbf{5.03} & \textbf{1.4.0} \\
		DTC \cite{Wang_DC-net2022} (AAAI) & 7(10\%) & 63(90\%) &       & 82.71 & 11.31 & 2.99 \\
		URPC \cite{Luo_efficient_semi-supervised} (MICCA) & 7(10\%) & 63(90\%) &       & 81.77 & 5.04  & 1.41 \\
		MC-Net \cite{Wu_Semi2021} (MICCAI) & 7(10\%) & 63(90\%) &       & 86.34 & 7.08  & 2.08 \\
		MC-Net+ \cite{Wu_Mutual_consistency2022} (MIA) & 7(10\%) & 63(90\%) &       & 87.10  & 6.68  & 2.00 \\
		Ours  & 7(10\%) & 63(90\%) &       & \textbf{87.83} & 6.05  & 1.45 \\
		UA-MT \cite{Yu_Uncertainty-aware2019} (MICCAI) & 14(20\%) & 56(80\%) &       & 85.87 & 5.06  & 1.54 \\
		SASSNet \cite{Li_Shape2020} (MICCAI) & 14(20\%) & 56(80\%) &       & 87.04 & 7.84  & 2.15 \\
		DTC \cite{Wang_DC-net2022} (AAAI) & 14(20\%) & 56(80\%) &       & 86.28 & 6.14  & 2.11 \\
		URPC \cite{Luo_efficient_semi-supervised} (MICCA) & 14(20\%) & 56(80\%) &       & 85.07 & 6.26  & 1.77 \\
		MC-Net \cite{Wu_Semi2021} (MICCAI) & 14(20\%) & 56(80\%) &       & 87.83 & \textbf{4.94} & 1.52 \\
		MC-Net+ \cite{Wu_Mutual_consistency2022} (MIA) & 14(20\%) & 56(80\%) &       & 88.51 & 5.35  & 1.54 \\
		Ours  & 14(20\%) & 56(80\%) &       & \textbf{88.52} & 5.44  & \textbf{1.37} \\
		\bottomrule
	\end{tabular}%
	\label{tab2}%
\end{table*}%
\begin{table}[htbp]
	\centering
	\caption{Model complexity comparison of multiple methods on the ACDC dataset.}
	\begin{tabular}{lll}
		\toprule
		\multirow{2}[4]{*}{Methods} & \multicolumn{2}{l}{Complexity} \\
		\cmidrule{2-3}          & Para.(M) & MACs(G) \\
		\midrule
		U-Net \cite{Ronneberger_UNet} & 2.21  & 2.99 \\
		UA-MT \cite{Yu_Uncertainty-aware2019} (MICCAI) & 2.21  & 2.99 \\
		SASSNet \cite{Li_Shape2020} (MICCAI) & 2.21  & 3.02 \\
		DTC \cite{Wang_DC-net2022} (AAAI) & 2.21  & 3.02 \\
		URPC \cite{Luo_efficient_semi-supervised} (MICCA) & 1.83  & 3.02 \\
		MC-Net \cite{Wu_Semi2021} (MICCAI) & 2.58  & 5.39 \\
		MC-Net+ \cite{Wu_Mutual_consistency2022} (MIA) & 2.21  & 2.99 \\
		Ours  & 2.39  & 3.31 \\
		\bottomrule
	\end{tabular}%
	\label{tab3}%
\end{table}%
\begin{table}[htbp]
	\centering
	\caption{Comparison of correct prediction rates of multiple prediction methods.}
	\begin{tabular}{ll}
		\toprule
		Methods & Accuracy \\
		\midrule
		Isensee et al. \cite{Isensee_Automatic_cardiac} & 0.92 \\
		Wolterink et al. \cite{Wolterink_Automatic_segmentation} & 0.86 \\
		Khened et al. \cite{Khened_Fully_convolutional} & 1.00 \\
		Ammar et al. \cite{Ammar_Automatic_cardiac} & 0.92 \\
		Ours  & \textbf{1.00} \\
		\bottomrule
	\end{tabular}%
	\label{tab4}%
\end{table}%
Table \ref{tab2} presents a quantitative comparison of various methods on the ACDC dataset, along with the fully supervised segmentation results of the base model U-Net. The method proposed in this paper achieves Dice, 95HD, and ASD scores of 87.83\%, 6.05 voxels, and 1.45 voxels, respectively, using only 10\% annotated data for training. The semi-supervised training with 10\% annotated data, compared to U-Net trained with 10\% annotated data, results in an improvement of 10.49 percentage points in Dice, a decrease of 3.13 voxels in 95HD, and a decrease of 1 voxel in ASD. Training the proposed method with 20\% annotated data yields results comparable to U-Net trained with all annotated data: Dice of 88.52\% compared to 91.65\%, 95HD of 5.44 voxels compared to 1.89 voxels, and ASD of 1.37 voxels compared to 0.56 voxels. These results indicate that the proposed semi-supervised segmentation method effectively utilizes unlabeled data and can achieve comparable performance to fully supervised training when the proportion of annotated data is low. Table \ref{tab3} presents a comparison of model complexity among various methods on the ACDC dataset.

Figure \ref{fig4} provides visual segmentation results for selected comparative methods. For segmentation results trained with 10\% annotated data, both MC-Net+ and the method proposed in this paper exhibit noticeable segmentation defects (highlighted by green arrows in Figure \ref{fig4}), with the method proposed in this paper showing significant improvement compared to MC-Net+. For segmentation results trained with 20\% annotated data, both MC-Net+ and the method proposed in this paper approximate manual annotation results closely. However, MC-Net+ still presents minor segmentation errors (highlighted by yellow arrows in Figure \ref{fig4}). In summary, it is evident that the segmentation results obtained using the method proposed in this paper are closer to the ground truth labels compared to other state-of-the-art algorithms. Particularly, when the proportion of annotated data is low, the proposed method demonstrates excellent segmentation performance, underscoring its outstanding semi-supervised segmentation capabilities.
\FloatBarrier
\subsection{Evaluation of the consistency of the calculation of clinical indicators}
\label{subsec3-3}
The paper conducts consistency evaluations between automatically extracted indices and manually determined indices using correlation analysis and Bland-Altman plots \cite{Bland_Statistical_methods}, as illustrated in Figures \ref{fig5} and \ref{fig6}. Figure \ref{fig5} demonstrates results obtained solely from semi-supervised training using seven cases (10\%) of annotated data, while Figure \ref{fig6} shows results obtained from fully supervised training using seventy cases (100\%) of annotated data. It is observed that almost all data points fall within the limits of agreement. For results obtained from semi-supervised training using only seven cases of annotated data, the volume calculation deviations for all three cardiac structures are within 10 ml, with deviations of 0.03\% and 0.02\% for left ventricular ejection fraction and right ventricular ejection fraction, respectively. For results obtained from fully supervised training using seventy cases of annotated data, the volume calculation deviations for all three cardiac structures are within 3 ml, with deviations of 0.01\% for both left ventricular ejection fraction and right ventricular ejection fraction. These findings indicate that the clinical indices calculated based on segmentation results in this paper exhibit good consistency with manually determined indices.

\subsection{Evaluation of disease diagnosis performance}
\label{subsec3-4}

The disease diagnosis classifier in this paper utilizes a two-layer ensemble classifier, and the confusion matrix for disease prediction on the test set of 50 cases is depicted in Figure \ref{fig7}. Figure \ref{fig7}(a) illustrates the prediction results of the first-layer classifier. It is evident that the first-layer classifier achieves precise classification for the NOR, HCM, and ARV classes, with a 100\% accuracy rate, but misclassifies MINF and DCM. This misclassification arises due to the shared symptom of reduced myocardial contraction function in both MINF and DCM. To further refine the prediction results, the paper incorporates a second-layer classifier specifically designed to distinguish between MINF and DCM. The prediction results of the second-layer classifier are shown in Figure \ref{fig7}(b), demonstrating excellent discrimination between MINF and DCM. Combining the predictions of the two-layer classifier, the classifier proposed in this paper achieves a 100\% accuracy rate on the test set of 50 cases. Table \ref{tab4} provides the prediction accuracy of various state-of-the-art methods on the ACDC test set, indicating the good predictive performance of the method proposed in this paper. Integrated with the semi-supervised automatic segmentation network proposed in this paper, the method exhibits low dependency on annotated data while maintaining excellent disease diagnosis performance.

\section{Conclusion}
\label{sec4}
The paper introduces a semi-supervised automatic segmentation and assisted diagnosis model, enabling fully automatic semi-supervised cardiac image segmentation, calculation of clinical physiological indices, and prediction of disease classification. The semi-supervised segmentation network integrates channel-wise prior convolutional attention and a complementary consistency semi-supervised segmentation architecture. This network possesses robust feature extraction capabilities and effectively utilizes unlabeled data from complementary information perspectives. The disease classifier adopts a dual-layer structure, incorporating a disease-specific classifier after the soft voting ensemble classifier to further enhance classification performance. Both the semi-supervised segmentation network and the disease classifier are benchmarked against other state-of-the-art methods, with experimental results showcasing excellent semi-supervised segmentation performance and high-accuracy classification effects of the proposed approach.
 
\section*{CRediT authorship contribution statement}
\textbf{Hejun Huang}: Conceptualization, Methodology, Formal analysis, Conducting experiments, Writing. \textbf{Zuguo Chen}: Data processing, Re-viewing and Editing, Funding acquisition. \textbf{Yi Huang}: Visualization, Supervision and Editing. \textbf{Guangqiang Luo}: Re-viewing and Editing. \textbf{Chaoyang Chen}: Supervision and Editing.  \textbf{Youzhi Song}: Data processing, Re-viewing and Editing.

\section*{Declaration of Competing Interes}
The authors declare that they have no known competing financial interests or personal relationships that could have appeared to influence the work reported in this paper.

\section*{Data availability}
Data will be made available on request.
\section*{Acknowledgments}
The research has been supported by the National Natural Science Foundation of China (62373144) and the Natural Science Foundation of Shenzhen City (JCYJ20210324101215039).








\bibliographystyle{cas-model2-names}

\bibliography{refs}



\end{document}